\documentclass[sigconf]{acmart}

\AtBeginDocument{%
  \providecommand\BibTeX{{%
    \normalfont B\kern-0.5em{\scshape i\kern-0.25em b}\kern-0.8em\TeX}}}

\setcopyright{none}
\acmConference[SIN 2020]{13th International Conference on Security of Information and Networks}{November 4--7, 2020}{Merkez, Turkey}
\acmBooktitle{13th International Conference on Security of Information and Networks (SIN 2020), November 4--7, 2020, Merkez, Turkey}
\acmPrice{15.00}
\acmDOI{10.1145/3433174.3433591}
\acmISBN{978-1-4503-8751-4/20/09}

\usepackage{amsmath}
\usepackage[linesnumbered,vlined,algoruled,boxed,lined]{algorithm2e}

\usepackage{pgfplots}
\pgfplotsset{compat=1.16}
\usepackage{pgfplotstable}
\usepackage{filecontents}

\usepackage{xcolor}
\usepackage{tikz}

\definecolor{bblue}{HTML}{4F81BD}
\definecolor{rred}{HTML}{C0504D}
\definecolor{ggreen}{HTML}{9BBB59}
\definecolor{ppurple}{HTML}{9F4C7C}

\begin{document}

\title{SpotCheck: On-Device Anomaly Detection for Android}


\author{Mark Vella}
\affiliation{%
  \institution{Dept. of Computer Science, University of Malta}
  \streetaddress{}
  \city{Msida}
  \country{Malta}}
\email{mark.vella@um.edu.mt}

\author{Christian Colombo}
\affiliation{%
\institution{Dept. of Computer Science, University of Malta}
  \streetaddress{}
  \city{Msida}
  \country{Malta}}
\email{christian.colombo@um.edu.mt}


\begin{abstract}
In recent years the PC has been replaced by mobile devices for many security sensitive operations, both from a privacy and a financial standpoint. While security mechanisms are deployed at various levels, these are frequently put under strain by previously unseen malware. An additional protection layer capable of novelty detection is therefore needed. In this work we propose SpotCheck, an anomaly detector intended to run on Android devices. It samples app executions and submits suspicious apps to more thorough processing by malware sandboxes. We compare Kernel Principal Component Analysis (KPCA) and Variational Autoencoders (VAE) on app execution representations based on the well-known system call traces, as well as a novel approach based on memory dumps. Results show that when using VAE, SpotCheck attains a level of effectiveness comparable to what has been previously achieved for network anomaly detection. Interestingly this is also true for the memory dump approach, relinquishing the need for continuous app monitoring.

\end{abstract}




\maketitle

\sloppy

\section{Introduction}
Mobile malware is an ever-increasing concern given the sensitive data and transactions nowadays stored and carried out on mobile devices, surpassing PC usage in many ways. Android is the leader in the mobile OS market \cite{statcounter2020}, and therefore the surge in malware targeting it in recent years comes as no surprise \cite{sophos2020}. Despite security considerations embedded within Android's design as well as anti-malware provisions at various stages of app deployment, limitations abound. Since the signature-based approach poses the main limitation with existing malware detection, an effective additional layer must provide for novelty detection \cite{kim2008detecting}. Anomaly detection builds a model of normal behavior by relying solely on a sufficiently large sample of benign apps. At runtime, those apps that deviate significantly from this model are flagged as suspicious, presenting possible malware. This contrasts with signature-based approaches that are devised to recognize known malware and their variants. Machine learning plays a central role through various clustering, classification and dimensionality reduction algorithms \cite{chandola2009anomaly}. In this work we consider two options: Kernel Principal Component Analysis (KPCA) and Variational Autoencoders (VAEs), for shallow and deep learning respectively \cite{an2015variational}, both previously experimented with for network anomaly detection.

\begin{figure}
\centering
\includegraphics[page=2,trim = 0mm 0mm 0mm 10mm, clip, width=0.9\linewidth]{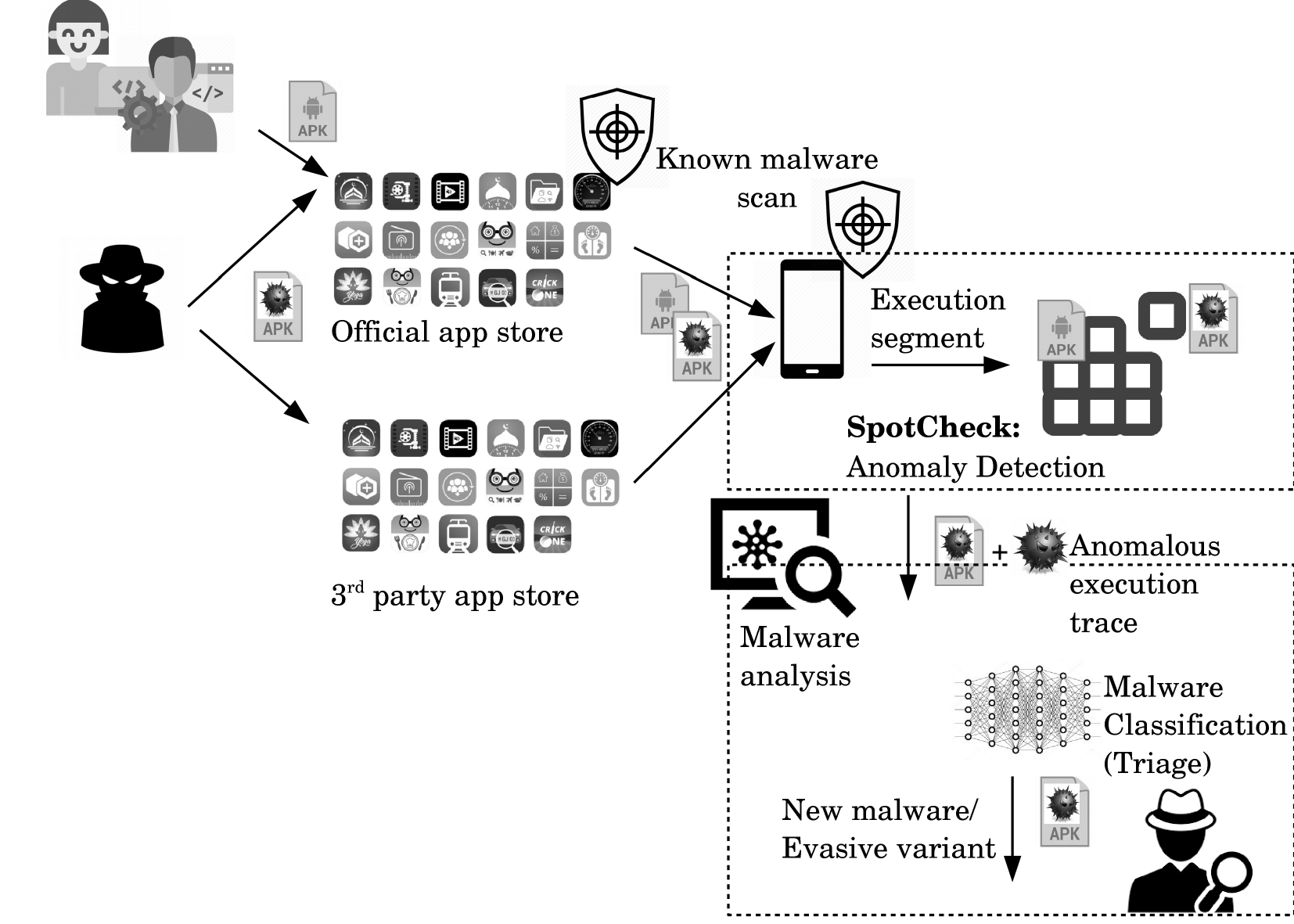}
\caption{SpotCheck for Android: on-device anomaly detection, submitting suspicious apps for further analysis.}
\label{fig:SpotCheck}
\end{figure}

As shown in Figure 1, SpotCheck is intended to operate on samples of on-device app execution segments, submitting apps with a sufficiently high anomaly score for deeper inspection by malware analysis. Rather than a standalone alert-raising monitor, SpotCheck acts as a precursor to malware triage. State-of-the-art malware analysis leverages machine learning to classify suspicious binaries according to malware families, with deep learning-based classifiers operating on system call traces being particularly effective \cite{hou2016deep4maldroid}, prior to manual analysis by experts. SpotCheck aims to benefit from machine learning in a similar manner, using dynamic analysis to capture app behavior in an obfuscation resilient way. The well-established system call trace representation of app behavior, as well as a more experimental process memory dump approach are taken into consideration. SpotCheck takes a sampling approach, conducting anomaly detection on execution segments. The net benefit of this precursor step to malware analysis is two-fold: first it can prioritize samples submitted for malware analysis; secondly, by providing the associated anomalous execution trace along with the app itself, malware analysis can be more focused. Overall, we make the following contributions:
\begin{itemize}
  \item We show that KPCA and VAE's effectiveness for Android anomaly detection is comparable to the use of VAE for network anomaly detection;
  \item We propose an experimental memory dump representation for app behavior, and which can be combined effectively with KPCA anomaly detection;
  \item Two datasets, comprising dynamic app behavior taken from Google Play and Virustotal, for both system call trace and process memory dump representations.
\end{itemize}

\section{Background}\label{section:background}

\subsection{Anomaly detection}
The core premise of malware anomaly detection is that malware should look and/or behave differently from benign apps \cite{chandola2009anomaly}. Therefore anomaly detection firstly has to model benign behavior, and secondly it needs some form of similarity measure from which to compute an anomaly score for the monitored apps. In proximity-based models malware is identified in terms of isolated datapoints, or else by forming its own clusters. Distance or density-based ones take a localized approach by considering only the closest points within a feature space, with malware expected to be excessively distant from the closest benign datapoint, or else located within a sparsely populated sub-space. These two approaches represent most of state-of-the-practice in network intrusion detection \cite{chio2018machine}.

Spectral methods combine dimensionality reduction with deviation-based anomaly detection. Whether using Principal Component Analysis (PCA) \cite{callegari2018improving} or Autoencoders (AE) \cite{an2015variational}, as computed/optimized from a benign-only dataset, a higher reconstruction error is expected for malware samples, therefore resulting in larger deviations. VAEs \cite{kingma2013auto} take a statistical approach and assume that datapoints are sampled from a specific probability distribution. A datapoint is anomalous if its probability falls below a certain threshold.

\subsection{Representing app behavior}
SpotCheck takes a dynamic analysis approach to represent app behavior. Obfuscating malicious intent from dynamic analysis is harder compared to static analysis. Capturing malware behavior as sequences of function/system calls is a well-established technique \cite{shankarapani2011malware}. An alternative approach is to analyze the residue of execution within process memory. That residue is made of the various data structures/objects that define the app state as a result of trace execution. Memory forensics \cite{case2017memory}, or the analysis of physical memory dumps, has received increased attention since the onset of advanced malware that does not leave any traces on disk. Yet this type of memory analysis is not suitable for non-rooted stock Android devices \cite{sylve2012acquisition}. Process-level memory dumps, on the other hand, are unencumbered by these restrictions. In fact most stock Android devices come equipped with a runtime that supports an extended version of HPROF memory dumps\footnote{https://developer.android.com/studio/profile/memory-profiler}. Yet, with this approach the timing of dump triggering becomes critical \cite{vella2018volatile}.

\section{SpotCheck's architecture}\label{section:spotcheck}

\subsection{Sampling app execution}
Since monitoring the entire app's execution is infeasible, we opt for sampling. The intuition is that when monitoring multiple runs, the sampling approach will eventually hit the sought after, discriminating, runtime behavior. We decide to explore both system call traces and process memory dumps to represent behavior. The prior approach serves as baseline, being well-established for security monitoring purposes. The latter is an experimental lightweight approach that avoids code instrumentation, yet it relies on identifying those discriminating data objects in-memory, and which may be short-lived. In this mode, the need to capture representative execution samples becomes even more critical.

\paragraph {System call traces.} Capturing Android app execution in terms of Linux system calls has been already widely explored for Android malware classification \cite{hou2016deep4maldroid}.  The chosen feature vector structure for the system call histogram representation is the 86-feature vector: 

\begin{equation*}\label{eq:syscallrep}
x \stackrel{def}{=}<\text{\it accept, access, bind, chdir, \ldots, writev}>
\end{equation*}

where each feature is a system call count, possibly spanning multiple processes for the same app, for some execution sample.

\paragraph{Process memory dumps.} The HPROF memory dump format provides an obvious choice in this case. Yet, choosing data objects with a high discriminating potential is not trivial. Unlike call traces there is no previous work to provide guidance. Among all Android and Java framework objects, we opt for those service classes returned by \texttt{android.content.Context.getSystemService()}. These classes act as interfaces to Android services accessible through the \texttt{android.os} namespace. As of its inception, the memory dump approach is bound to be more limited than the comprehensive system call tracing approach, given it has to rely solely on in-memory residue of execution. The chosen structure for the HPROF histogram representation is the 72-feature vector:
\begin{equation*}\label{eq:memrep}
x \stackrel{def}{=}<\text{\it AccessibilityManager, \ldots, WindowManager}>
\end{equation*}

where once again each feature is for individual apps, possibly spanning multiple processes. 

In both representations features are scaled using an Attribute Ratio method that normalizes counts as a fraction of the total counts per vector: $\hat{x}\stackrel{def}{=}<a_{i}/||x||_{1}, ..., a_{n}/||x||_{1}>$. The normalized total of counts is therefore 1 per datapoint ($||\hat{x}||_{1}=1$), offsetting irregularities derived from sampling executions of different lengths.

\subsection{Kernel Principal Component Analysis (KPCA) for anomaly detection}

KPCA is a non-linear variant of classic PCA. Like its linear counterpart it performs dimensionality reduction in a way that maximises information retention, expressed in terms of variance. Either eigen or singular value decomposition can be used to map from the original $n$-dimensional feature space to a latent $r$-dimensional one. By setting $r$ to 2 or 3, it is possible to visualise high-dimensional datasets. Sticking to eigendecomposition (slower for PCA, but the only option for efficient KPCA), given a (centered) $n$-dimensional dataset $X_{n}$, the covariance matrix $X^{T}.X$ is first computed, followed by its eigendecomposition to $W.\lambda.W^{-1}$. The columns in $W$ store the orthogonal eigenvectors, indicating directions of most variance. $\lambda$ is a diagonal matrix of eigenvalues. $W_{r}$ is the result of reordering $W$ according to $\lambda$, with the columns associated with the largest eigenvalue ordered left-most, and subsequently dropping all but the first $r$ columns. Latent space mapping is computed as $Z_{r}=W_{r}.X$, while the inverse transform comprises $X=Z_{r}.W_{r}^{T}$. Note that $W_{r}^{T}.W_{r}$ is the identity matrix only when $r=n$, and therefore constitutes a lossy transform otherwise.

KPCA provides non-linearity by means of kernel methods by mapping $X_{n}$ to $X_{m}=\phi(X_{n})$, where $m>n$. The kernel trick is used to maintain tractable computation in higher dimensions by using a kernel function $k(X)=\phi(X^{T}).\phi(X)$. If it can be assumed that the higher-dimensional space follows a Gaussian distribution, the radial basis function (rbf) kernel is used:

\begin{equation*}\label{eq:rbf}
k(x,y) \stackrel{def}{=}e^{-\gamma||x-y||^{2}},\text{where } \gamma = \frac{1}{2\sigma^{2}} > 0
\end{equation*}

with $\gamma$ presenting a learnable parameter corresponding to a training dataset $X$. An optimal $\gamma$ is typically computed using a grid search with the mean squared error (MSE) used as the reconstruction error. The premise for using KPCA for anomaly detection is that, during testing, data points $x^{(i)}$ form a different distribution than one from which the training dataset is derived, returning a higher reconstruction error.

\SetKwData{Ben}{Benign Apps}
\SetKwData{Test}{Monitored Apps}
\SetKwData{SmpMode}{Mode}
\SetKwData{Recon}{Anomaly\_Scores}
\SetKwData{Tresh}{Threshold}
\SetKwData{ReconErr}{ReconErr}
\SetKwData{Trans}{KPCA\_Transform}
\SetKwData{Anom}{Anomaly}
\SetKwData{Norm}{Normal}
\SetKwFunction{GS}{Grid\_Search}
\SetKwFunction{ED}{Eigendecomposition}
\begin{algorithm}[!t]
\footnotesize
 \KwIn{ \SmpMode [SysCall trace $|$ HPROF dump], \Tresh $\alpha$,\\\Ben $X$, \Test $x^{(1)}, ..., x^{(N)} \in X'$}
 \KwOut{ $MSE(x^{(i)},\hat{x}^{(i)})$ \Recon[]} 
 \BlankLine
 $W_{r}, W_{r}^{T} \leftarrow$ \ED($X$)\\
 $\gamma \leftarrow$ \GS($X$)\\
 \Recon[] = \{\}
 \BlankLine
  \For{$i\gets1$ \KwTo $N$ }{
    $z^{(i)}\gets \Trans(x^{(i)},W_{r},\gamma)$\\
    $\hat{x}^{(i)}\gets \Trans^{-1}(z^{(i)},W_{r}^{T})$\\
    \ReconErr$=MSE(x^{(i)},\hat{x}^{(i)})$\\
    \BlankLine
     \eIf{(\ReconErr$ > \alpha$)}{
         \Recon[]$\gets (x^{(i)},\ReconErr,\Anom)$
      }{
        \Recon[]$\gets (x^{(i)},\ReconErr,\Norm)$
      } 
  }
 \BlankLine
 \Return{\Recon[]}

\caption{SpotCheck's KPCA anomaly detector}\label{alg:spotcheckpca}
\end{algorithm}
\normalsize

The KPCA anomaly detector is shown in Algorithm \ref{alg:spotcheckpca}. The training dataset $X$, composed solely of benign apps, is used for computing $W_{r}$ and $W_{r}^{T}$, as well as searching for optimal $\gamma$ (lines 1-2). For each monitored app $x^{(i)} \in X'$, its latent representation in $r$-dimensions, $z^{(i)}$, is computed and subsequently recovered as $\hat{x}^{(i)}$ (lines 5-6). Due to the lossy transforms involved, $x^{(i)} \neq \hat{x}^{(i)}$ and their mean squared error (MSE) is taken as the reconstruction error (line 7). Whenever this error exceeds a threshold $\alpha$, $x^{(i)}$ is flagged as anomalous (lines 8-11).

\subsection{Anomaly detection with Variational Autoencoder (VAE) for anomaly detection}

\begin{figure}
\centering
\includegraphics[page=3,trim = 5mm 60mm 5mm 15mm, clip, width=1\linewidth]{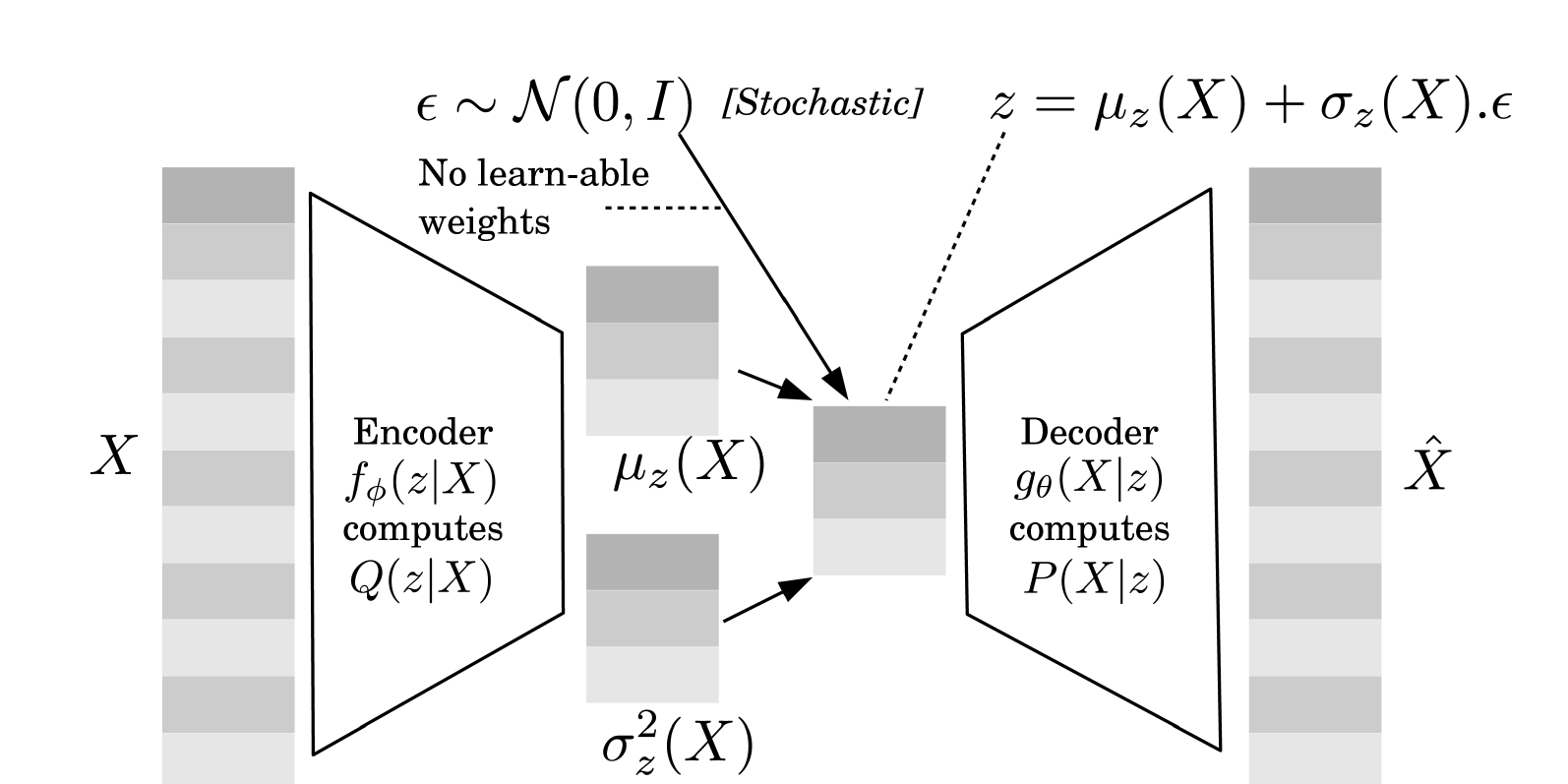}
\caption{ VAE topology: an encoder followed by a stochastic decoder, optimized wrt reconstruction probability $P(X)$.}
\label{fig:VAE}
\end{figure}

VAEs \cite{kingma2013auto} approximate a probability distribution $P(X)$ to fit a data sample $X$ using neural networks as shown in Figure \ref{fig:VAE}. The decoder network $g_{\theta}(X|z)$ learns to \emph{generate datapoints similar to $X$} using a prior distribution defined over a much simpler latent space $P(z)$, namely the standard isotropic Gaussian $\mathcal{N}(0,I)$. In the latent space, datapoints $z$ have reduced and independent dimensions. The original feature space is assumed to follow a multivariate Gaussian (for continuous data) or Bernoulli (for binary) distributions. The complex relationship between the latent and original spaces is captured by $g_{\theta}(X|z)$ \cite{doersch2016tutorial}.

VAEs maximize $P(X)$, the \emph{reconstruction probability} i.e. the probability of computing a distribution that is most likely to in turn produce $X$, and which is exactly what renders VAEs suitable for anomaly detection. This way $g()$ is bound to produce outputs similar to $X$ seen during training, but not to inputs taken from different distributions. $g()$'s output, denoted by $\hat{X}$, represents a generated datapoint for a given $z$. Whenever considering a single datapoint, $\hat{X}$ is taken as the mean. With (multivariate) Gaussians, however, one needs to also consider $\sigma^{2}$, the covariance. This can either be represented as an additional output to $\mu=\hat{X}$ \cite{kingma2013auto}, or else taken to be a fixed hyperparameter \cite{doersch2016tutorial}. The role of the encoder $f_{\phi}(z|X)$ is to compute $Q(z|X)$ in a way that is as close as possible to $P(z|X)$, or the probability distribution in the latent space that is most likely to reproduce $X$. An intermediate function $z=h_{\phi}(\epsilon,X)$ is used for sampling datapoints $z$ in the latent space. $f_{\phi}(z|X)$ computes $\mu_{z}(X)$ and $\sigma_{z}^{2}(X)$, the mean and covariance of the latent space respectively. Here $h_{\phi}(\epsilon,X) \stackrel{def}{=}\mu_{z}(X)+\sigma_{z}(X).\epsilon$ and $\epsilon \sim \mathcal{N}(0,I)$. In this manner no learnable weight is associated with a stochastic node, and backpropagation can proceed as usual. The resulting loss function is the negative of the ELBO objective function:

\begin{equation*}\label{eq:elbo}
 -ve\;ELBO \stackrel{def}{=}\mathcal{D}_{KL}(Q(z|x^{(i)})||\mathcal{N}(0,I)) - \mathbf{E}_{Q(z|x^{(i)})}[log\;P(x^{(i)}|z)] 
\end{equation*}

and which is defined in a way to keep $log\;P(X)$ close to 0, over choices for $\phi,\theta$. The first term on the right hand-side penalizes any encodings produced by $Q(z|X)$ not following the assumed simple latent distribution. The second term is the reconstruction error.

\SetKwData{Ben}{Benign Apps}
\SetKwData{Test}{Monitored Apps}
\SetKwData{SmpMode}{Mode}
\SetKwData{Recon}{Anomaly\_Scores}
\SetKwData{Tresh}{Threshold}
\SetKwData{ReconProb}{ReconProb}
\SetKwData{Anom}{Anomaly}
\SetKwData{Norm}{Normal}
\SetKwFunction{MINBATCH}{Mini\_Batch}
\begin{algorithm}[!t]
\footnotesize
 \KwIn{ \SmpMode [SysCall trace $|$ HPROF dump], \Tresh $\alpha$,\\\Ben $X$, \Test $x^{(1)}, ..., x^{(N)} \in X'$}
 \KwOut{ $P(x^{(i)})$ \Recon[]} 
 \BlankLine
 $\phi,\theta \leftarrow$ \MINBATCH($X$)\\
 \Recon[] = \{\}
 \BlankLine
  \For{$i\gets1$ \KwTo $N$ }{
    $\mu_{z^{(i)}},\sigma^{2}_{z^{(i)}} \leftarrow f_{\phi}(z|x^{(i)})$\\
    \BlankLine
    \For{$l\gets1$ \KwTo $L$ }{
        $z^{(i,l)}\sim \mathcal{N}(\mu_{z^{(i)}},\sigma^{2}_{z^{(i)}} )$\\
        $\mu_{\hat{x}^{(i,l)}},\sigma^{2}_{\hat{x}^{(i,l)}}\gets g_{\theta}(\hat{x}^{(i,l)}|z^{(i,l)})$
    }
    \ReconProb = $P(x^{(i)})\gets \frac{1}{L}\sum\limits_{l=1}^LP(x^{(i,l)};\mu_{\hat{x}^{(i,l)}},\sigma^{2}_{\hat{x}^{(i,l)}})$\\
    
     \eIf{(\ReconProb $< \alpha$)}{
         \Recon[]$\gets (x^{(i)},\ReconProb,\Anom)$
      }{
        \Recon[]$\gets (x^{(i)},\ReconProb,\Norm)$
      } 
  }
 \BlankLine
 \Return{\Recon[]}

\caption{SpotCheck's VAE anomaly detector}\label{alg:spotcheckvae}
\end{algorithm}
\normalsize
SpotCheck's VAE anomaly scores are based on datapoint-wise reconstruction probabilities $P(x^{(i)})$, as shown in Algorithm \ref{alg:spotcheckvae} (lines 9-12), in turn hinging on the learned $\phi, \theta$ (line 1). We take two approaches for dealing with the computed covariance $\sigma^{2}$ at the feature space: \emph{a)} as a learned layer, or as a \emph{b)} hyperparameter fixed at 1. In both cases we assume a Gaussian distribution since we are dealing with scaled frequencies in the 0-1 range. In the first case, for $L=1$, $-\mathbf{E}_{Q(z|x^{(i)})}[log\;P(x^{(i)}|z)]$ becomes \cite{nix1994estimating}:
\begin{equation*}\label{eq:nll}
NLL_{Gaussian} = \sum\limits_{i}\frac{log\;\sigma_{\hat{x}^{(i)}}^{2}}{2}+\frac{(x^{(i)}-\mu_{\hat{x}^{(i)}})^{2}}{2\sigma_{\hat{x}^{(i)}}^{2}}
\end{equation*}

In the second case, fixing $\sigma^{2}=1$ renders the terms with $\sigma^{2}$ constant, and which reduces to the commonly-used mean squared error (MSE). The KL divergence term has the closed-form \cite{kingma2013auto}:

\begin{equation*}\label{eq:kl}
\mathcal{D}_{KL}(Q(z|x^{(i)})||\mathcal{N}(0,I)) = \frac{1}{2}\sum\limits_{i}( 1 + log\;(\sigma^{2}_{z^{(i)}}) - \mu^{2}_{z^{(i)}} - \sigma^{2}_{z^{(i)}} )
\end{equation*}

 We opt for the Adam optimizer, with $L=1$ as originally suggested \cite{kingma2013auto}. ReLU activation is used for all layers except for $\sigma^{2}(\hat{X})$, which uses linear activation as originally suggested \cite{kingma2013auto}, with a bias term of $1\times10^{-4}$ to avoid a divide by zero when computing $NLL_{Gaussian}$. $\mu(\hat{X})$ uses sigmoid activation followed by feature scaling to match input feature scaling.

Lines 3-12 take the trained VAE and a set of input traces/dumps in order to compute anomaly scores. For each $x^{(i)}$, the latent space $\mu_{z^{(i)}},\sigma^{2}_{z^{(i)}}$ vectors are computed (line 4) and then used to sample $L$ $z^{(i,l)}$ points in latent space directly from $\mathcal{N}(\mu_{z^{(i)}},\sigma^{2}_{z^{(i)}} )$ (line 6). We set $L=128$ in order to match the training batch size. The feature space distribution parameters are taken as the mean of all predicted individual datapoints (lines 7 and 8).

\section{Experimentation}\label{section:eval}

SpotCheck experimentation compares KPCA with VAE across the two chosen representations.

\subsection{Datasets}

 A total of 3K apps were used: 2K benign apps downloaded from Google Play, and 1K malicious apps obtained from VirusTotal\footnote{https://www.virustotal.com/}. Two datasets\footnote{Available at https://github.com/mmarrkv/spotcheck\_ds}, one for each representation type, are a result of: Component traversal, as suggested in related work \cite{hou2016deep4maldroid}, to maximize runtime behavior coverage; and subsequently, a total of 200 (repeatable) pseudo-random UI events. An Android Pie image (API level 28) was used.  System call tracing was implemented with frida-server 12.10.4. Eclipse MAT 1.10/calcite v1.4 plugin was used for HPROF parsing.
 
Figure \ref{fig:datahistograms} visualizes the dataset features for both representations in terms of mean (scaled) frequencies per system call/service class, and for both benign apps and malware. The  system call histogram are characterized by a few dominant calls. In each case the three most frequent calls are \texttt{write}, \texttt{read}, and \texttt{ioctl}; and which correspond to input/output/ipc respectively, with \texttt{write} being particularly more frequent for malware than benign. \texttt{gettimeofday} and \texttt{recvfrom} are more frequent in benign apps. On the other hand \texttt{close} and \texttt{writev} rank higher for malware. \texttt{mmap} and \texttt{munmap} rank high in both cases, but even more so in malware.


In the case of benign HPROF dumps there are no dominating attributes, with the most frequent service class instances corresponding to \texttt{AudioManager}, \texttt{DisplayManager}, \texttt{TelephonyManager} and \texttt{UserManager}. On the contrary, \texttt{TelephonyManager} dominates for malware apps, and which more than doubles the benign app frequency. The number of \texttt{AccessibilityManager} instances are also doubled, although not being as dominant as the previous class. Other system classes with a high frequency for malware are: \texttt{AlarmManager}, \texttt{AudioManager}, \texttt{ConnectivityManager}, \texttt{DisplayManager}, \texttt{InputMethodManager} and \texttt{SubscriptionManager}.



\pgfplotstableread[col sep=comma]{testdata1.csv}\datatable
\makeatletter
\pgfplotsset{
    /pgfplots/flexible xticklabels from table/.code n args={3}{%
        \pgfplotstableread[#3]{#1}\coordinate@table
        \pgfplotstablegetcolumn{#2}\of{\coordinate@table}\to\pgfplots@xticklabels
        \let\pgfplots@xticklabel=\pgfplots@user@ticklabel@list@x
    }
}
\makeatother

\begin{figure}
\centering
\begin{tabular}{cc}

\begin{tikzpicture}[scale=0.55]
\begin{axis}[
height = 6cm,
ymajorgrids = true,
major x tick style = transparent,
ybar, bar width=3pt, ymin=0, ymax=0.6,
xlabel=Linux system call,
ylabel=Benign trace freq.,
flexible xticklabels from table={testdata1.csv}{syscall}{col sep=comma},
xticklabel, 
xtick=data,
nodes near coords align={vertical}]
\addplot[ggreen,fill] table[x expr=\coordindex,y=mean]{\datatable};

\end{axis}

\end{tikzpicture}

&

\pgfplotstableread[col sep=comma]{testdata2.csv}\datatable
\makeatletter
\pgfplotsset{
    /pgfplots/flexible xticklabels from table/.code n args={3}{%
        \pgfplotstableread[#3]{#1}\coordinate@table
        \pgfplotstablegetcolumn{#2}\of{\coordinate@table}\to\pgfplots@xticklabels
        \let\pgfplots@xticklabel=\pgfplots@user@ticklabel@list@x
    }
}
\makeatother

\begin{tikzpicture}[scale=0.55]
\begin{axis}[
height = 6cm,
ymajorgrids = true,
major x tick style = transparent,
ybar, bar width=3pt, ymin=0, ymax=0.6,
xlabel=Linux system call,
ylabel=Malicious trace freq.,
flexible xticklabels from table={testdata2.csv}{syscall}{col sep=comma},
xticklabel, 
nodes near coords align={vertical}]
\addplot[rred,fill] table[x expr=\coordindex,y=mean]{\datatable};

\end{axis}

\end{tikzpicture}

\\

\pgfplotstableread[col sep=comma]{testdata3.csv}\datatable
\makeatletter
\pgfplotsset{
    /pgfplots/flexible xticklabels from table/.code n args={3}{%
        \pgfplotstableread[#3]{#1}\coordinate@table
        \pgfplotstablegetcolumn{#2}\of{\coordinate@table}\to\pgfplots@xticklabels
        \let\pgfplots@xticklabel=\pgfplots@user@ticklabel@list@x
    }
}
\makeatother

\begin{tikzpicture}[scale=0.55]
\begin{axis}[
height = 6cm,
ybar, bar width=3pt, ymin=0, ymax=0.2,
ymajorgrids = true,
major x tick style = transparent,
xlabel=Android system service class,
ylabel=Benign instance freq.,
flexible xticklabels from table={testdata3.csv}{hprofclass}{col sep=comma},
xticklabel, 
nodes near coords align={vertical}]
\addplot[ggreen,fill] table[x expr=\coordindex,y=mean]{\datatable};

\end{axis}

\end{tikzpicture}

&

\pgfplotstableread[col sep=comma]{testdata4.csv}\datatable
\makeatletter
\pgfplotsset{
    /pgfplots/flexible xticklabels from table/.code n args={3}{%
        \pgfplotstableread[#3]{#1}\coordinate@table
        \pgfplotstablegetcolumn{#2}\of{\coordinate@table}\to\pgfplots@xticklabels
        \let\pgfplots@xticklabel=\pgfplots@user@ticklabel@list@x
    }
}
\makeatother

\begin{tikzpicture}[scale=0.55]
\begin{axis}[
height = 6cm,
ybar, bar width=3pt, ymin=0, ymax=0.2,
ymajorgrids = true,
major x tick style = transparent,
xlabel=Android system service class,
ylabel=Malicious instance freq.,
flexible xticklabels from table={testdata1.csv}{syscall}{col sep=comma},
xticklabel, 
nodes near coords align={vertical}]
\addplot[rred,fill] table[x expr=\coordindex,y=mean]{\datatable};

\end{axis}

\end{tikzpicture}


\end{tabular}
\caption{Scaled mean frequencies for system call trace (top) and process memory dump (bottom) features, for benign (left) and malware (right) apps.}
\label{fig:datahistograms}
\end{figure}
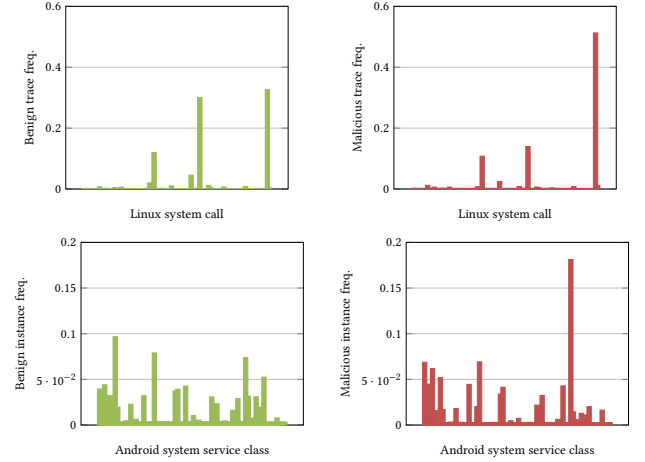

\subsection{Results}

\begin{figure*}
\centering
\begin{tabular}{cc}

\begin{tikzpicture}[scale=0.6]
    \begin{axis}[
        width  = 0.7*\textwidth,
        height = 7cm,
        major x tick style = transparent,
        ybar=0.5*\pgflinewidth,
        bar width=4pt,
        ymajorgrids = true,
        ylabel = {Accuracy metric},
        xlabel = {Anomaly detector (for system calls)},
        symbolic x coords={KPCA, VAE-1,VAE-2,VAE-3,VAE-4,VAE-5,VAE-6},
        xtick = data,
        scaled y ticks = false,
        enlarge x limits=0.1,
        ymin=0,
        legend cell align=left,
        legend style={
                at={(1.0,0.9)},
                anchor=south east,
                column sep=1ex
        }
    ]
        \addplot[style={bblue,fill=bblue,mark=none}]
            coordinates {(KPCA,0.7080028538) (VAE-1,0.6949519176) (VAE-2,0.6951163667) (VAE-3,0.6937742088) (VAE-4, 0.6950961268) (VAE-5,0.6917616044) (VAE-6,0.6935857248) };

        \addplot[style={rred,fill=rred,mark=none}]
             coordinates {(KPCA,0.8641584158) (VAE-1,0.5125858124) (VAE-2,0.5085158151) (VAE-3,0.5097087379) (VAE-4, 0.5122235157) (VAE-5,0.5085995086) (VAE-6,0.5097560976) };

        \addplot[style={ggreen,fill=ggreen,mark=none}]
             coordinates {(KPCA,0.9909173479) (VAE-1,0.6239554318) (VAE-2,0.5821727019) (VAE-3,0.5849582173) (VAE-4, 0.6128133705) (VAE-5,0.5766016713) (VAE-6,0.5821727019) };
        
        \addplot[style={ppurple,fill=ppurple,mark=none}]
            coordinates { (KPCA,0.7661516854) (VAE-1,0.4349514563) (VAE-2,0.4514038877) (VAE-3,0.4516129032) (VAE-4, 0.44) (VAE-5,0.4549450549) (VAE-6,0.453362256) };

        \legend{AUC ROC,F1 score,Recall,Precision}
    \end{axis}
\end{tikzpicture}

&

\begin{tikzpicture}[scale=0.6]
    \begin{axis}[
        width  = 0.7*\textwidth,
        height = 7cm,
        major x tick style = transparent,
        ybar=0.5*\pgflinewidth,
        bar width=4pt,
        ymajorgrids = true,
        ylabel = {Accuracy metric},
        xlabel = {Anomaly detector (for memory dumps)},
        symbolic x coords={,KPCA, VAE-1,VAE-2,VAE-3,VAE-4,VAE-5,VAE-6},
        xtick = data,
        scaled y ticks = false,
        enlarge x limits=0.1,
        ymin=0,
        legend cell align=left,
        legend style={
                at={(1,0.9)},
                anchor=south east,
                column sep=1ex
        }
    ]
        \addplot[style={bblue,fill=bblue,mark=none}]
            coordinates {(KPCA,0.6990056918) (VAE-1,0.7165016346) (VAE-2,0.7141202343) (VAE-3,0.712491474) (VAE-4, 0.6946677211) (VAE-5,0.699852412) (VAE-6,0.5611769998) };

        \addplot[style={rred,fill=rred,mark=none}]
             coordinates {(KPCA,0.8784965035) (VAE-1,0.5115830116) (VAE-2,0.5050691244) (VAE-3,0.5043478261) (VAE-4, 0.4962025316) (VAE-5,0.5089201878) (VAE-6,0.4224945926) };

        \addplot[style={ggreen,fill=ggreen,mark=none}]
             coordinates {(KPCA,0.9691417551) (VAE-1,0.8079268293) (VAE-2,0.8353658537) (VAE-3,0.8841463415) (VAE-4, 0.8963414634) (VAE-5,0.8262195122) (VAE-6,0.8932926829)};

        \addplot[style={ppurple,fill=ppurple,mark=none}]
             coordinates {(KPCA,0.8033573141) (VAE-1,0.3742937853) (VAE-2,0.3619550859) (VAE-3,0.3527980535) (VAE-4, 0.3430571762) (VAE-5,0.3677069199) (VAE-6,0.2766761095) };

        \legend{AUC ROC,F1 score,Recall,Precision}
    \end{axis}
\end{tikzpicture}

\end{tabular}
\caption{Classification accuracy for system call traces (left) and memory dumps (right).}
\label{fig:classification}
\end{figure*}
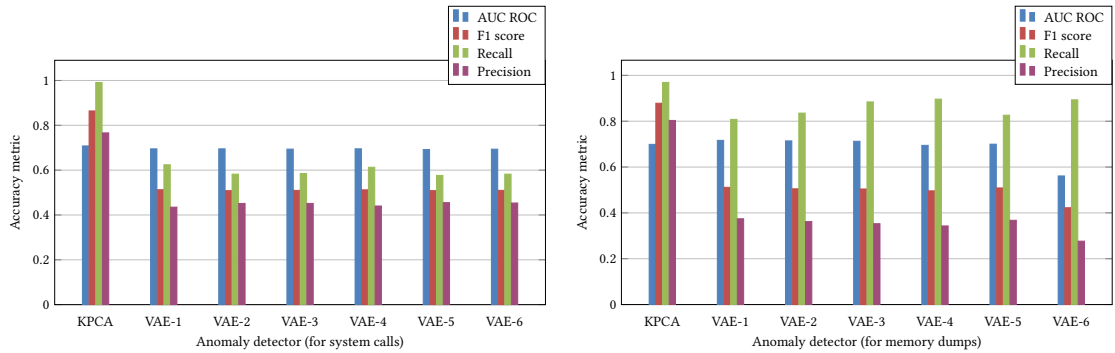

Figure \ref{fig:classification} shows a comparison of the classification accuracy obtained for KPCA and the VAE configurations, across both app execution representations. The KPCA implementation uses the RBF kernel in order to match the VAEs Gaussian approximation. The Grid Search for $\gamma$ uses 3-fold cross-validation in the 0.01-0.5 range. A two dimensional latent space is adopted for visualisation benefits. For VAE we try out 6 configurations in total. Configurations 1-3 use the negative log likelihood for Gaussian (NLL) in the loss function, and 50-25, 50-35-25 and 50-25-2 topologies respectively. Configurations 4-6 follow the same order, but this time making use of the MSE reconstruction error. In all cases 2,000 epochs was sufficient for loss function convergence. A 70/15/15 train/validation/test split is used for the benign datasets. Given the anomaly detection context, the malware datasets were only used during testing.

\begin{figure*}
\centering
\begin{tabular}{ccc}

\includegraphics[trim = 20mm 20mm 20mm 20mm, clip, 
width=0.24\linewidth]{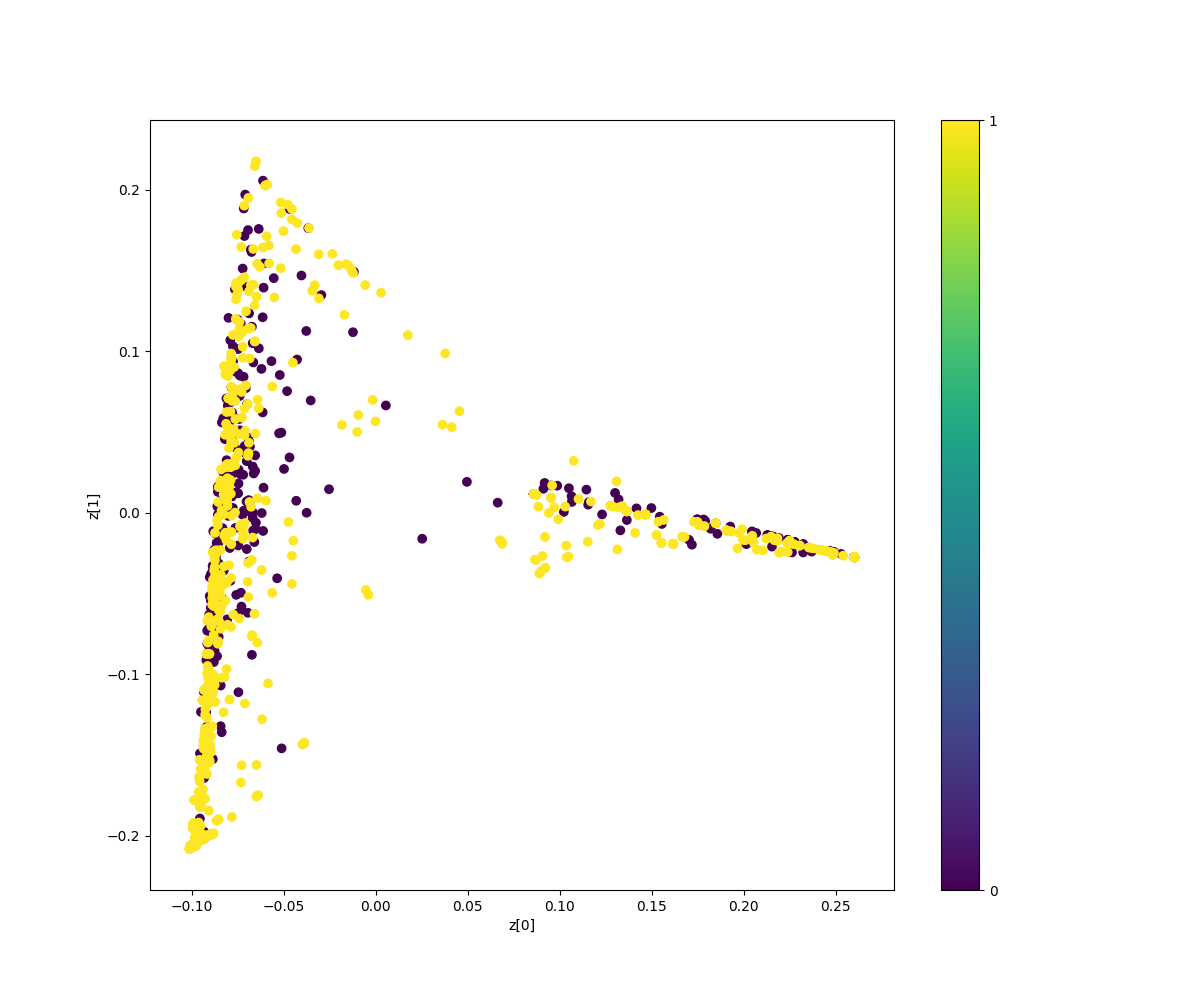}
&

\includegraphics[trim = 20mm 20mm 20mm 20mm, clip, 
width=0.24\linewidth]{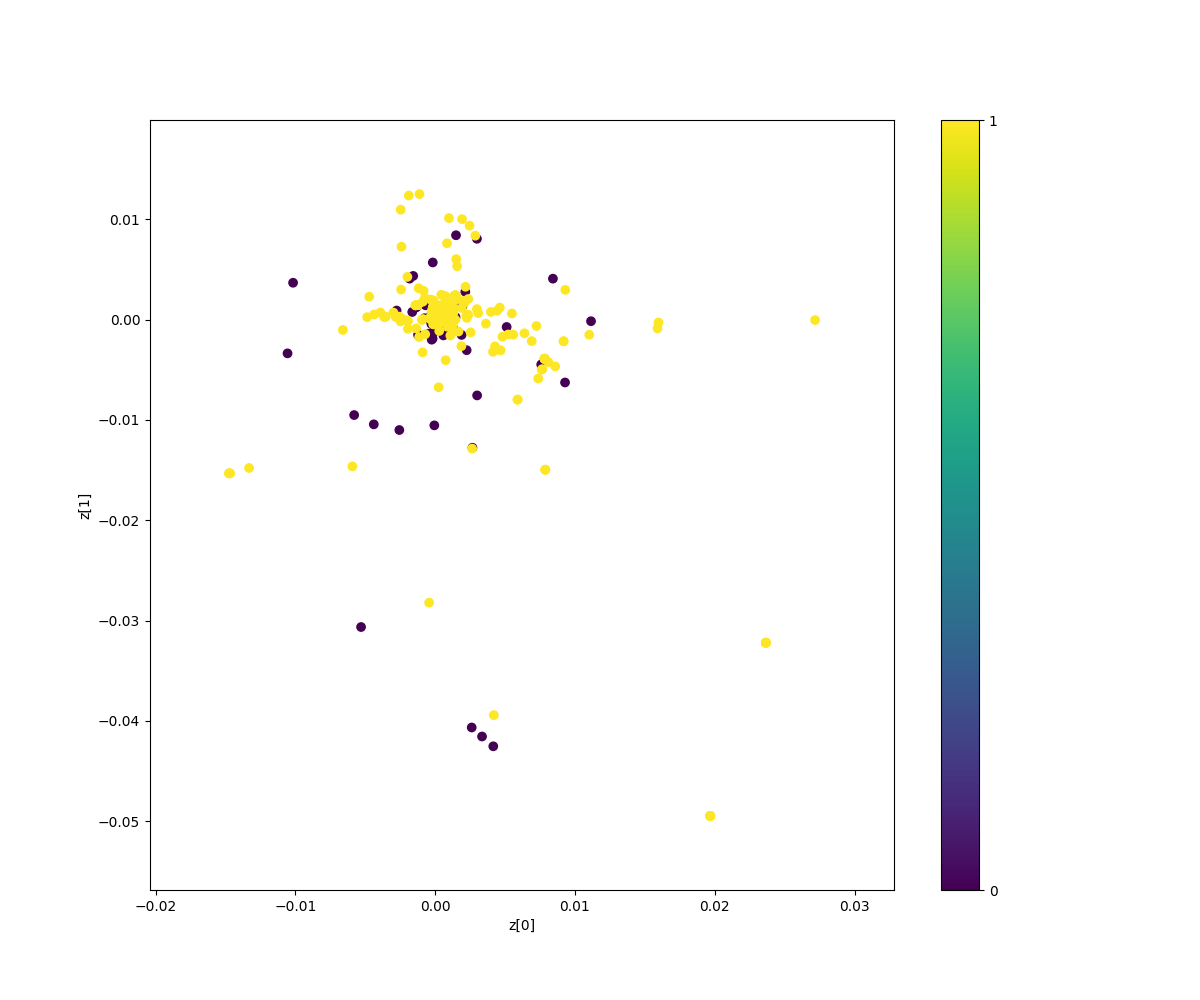}
&

\includegraphics[trim = 20mm 20mm 20mm 20mm, clip, 
width=0.24\linewidth]{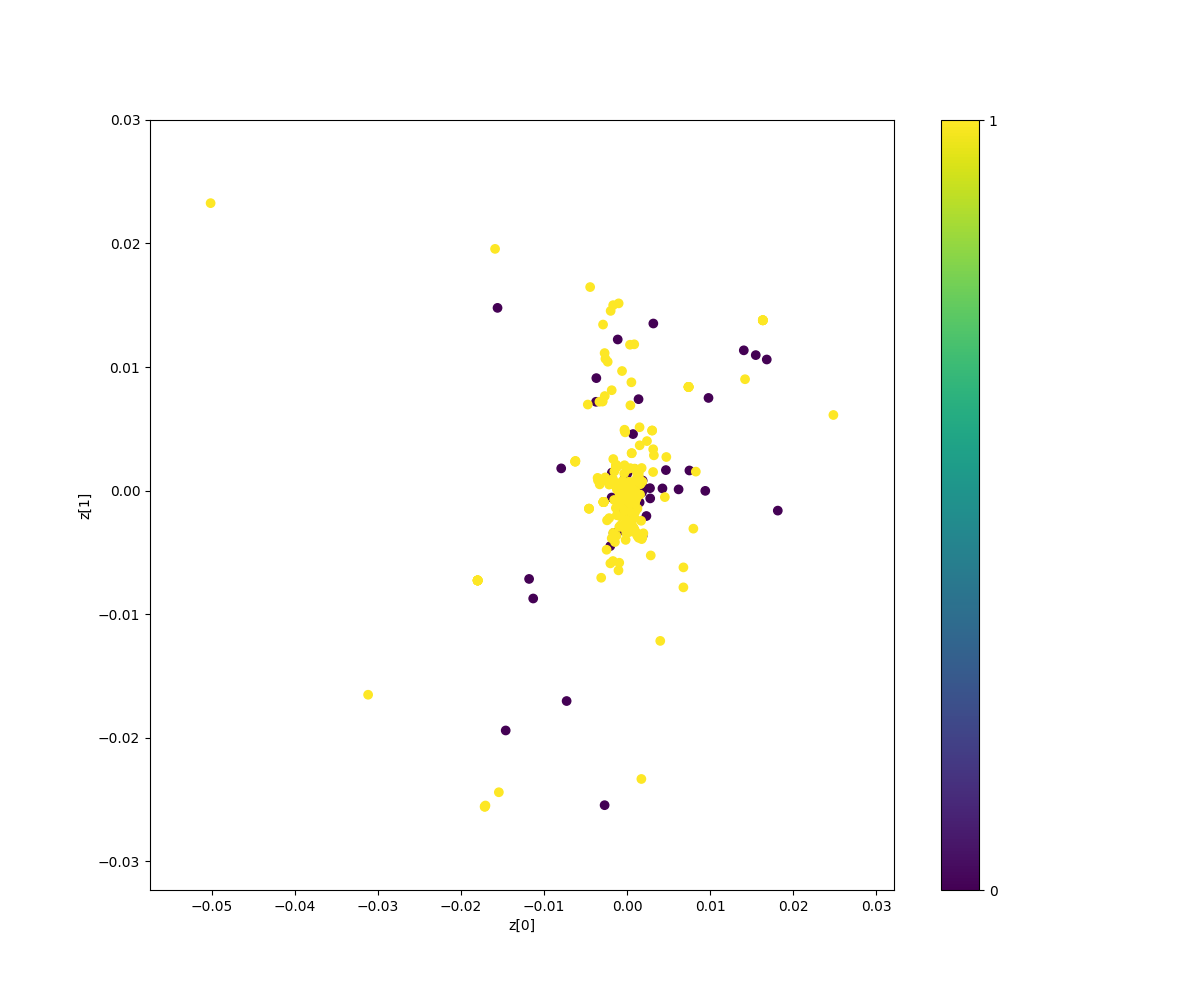}

\\

\includegraphics[trim = 20mm 20mm 20mm 20mm, clip, width=0.24\linewidth]{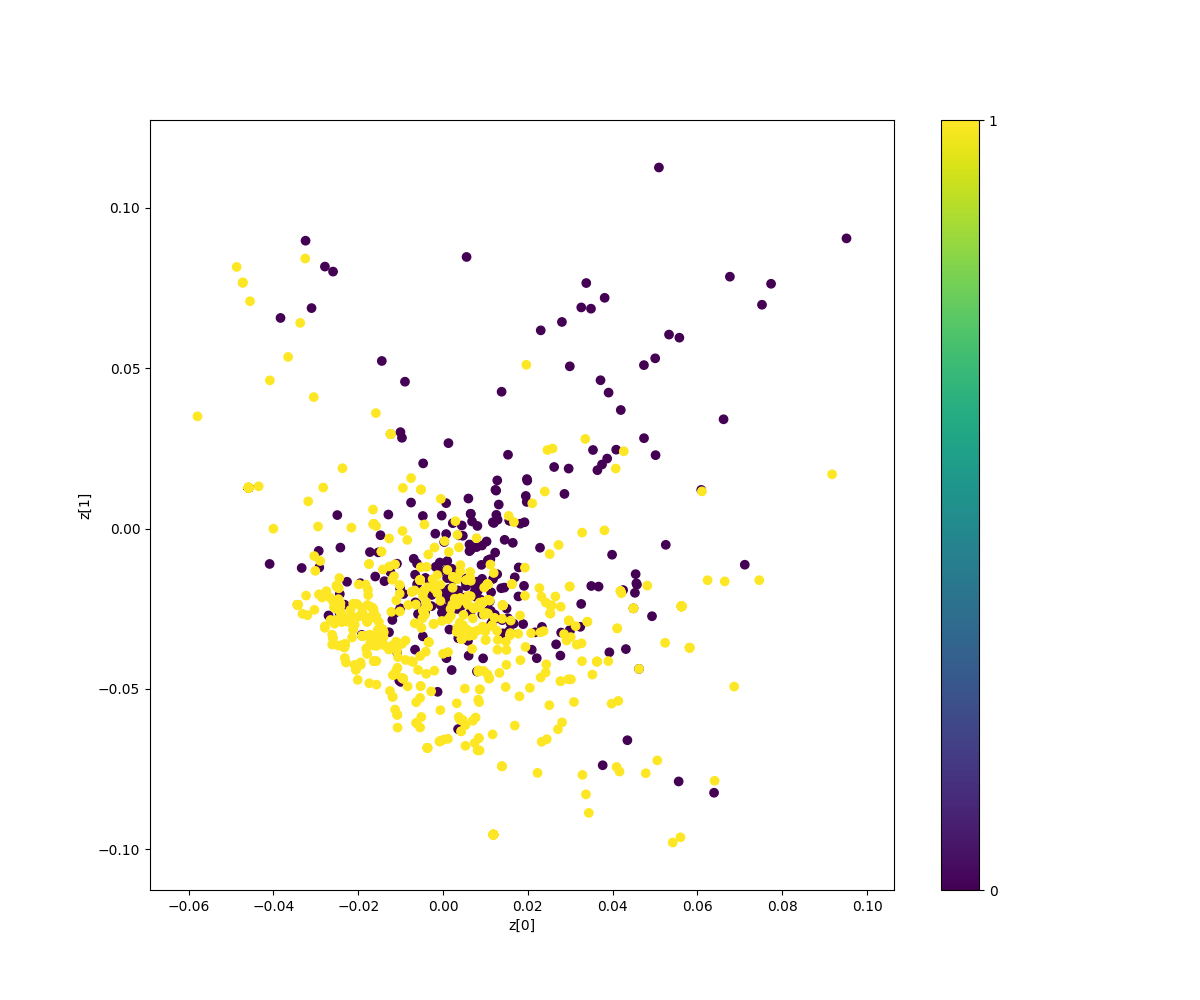}

&

\includegraphics[trim = 20mm 20mm 20mm 20mm, clip, width=0.24\linewidth]{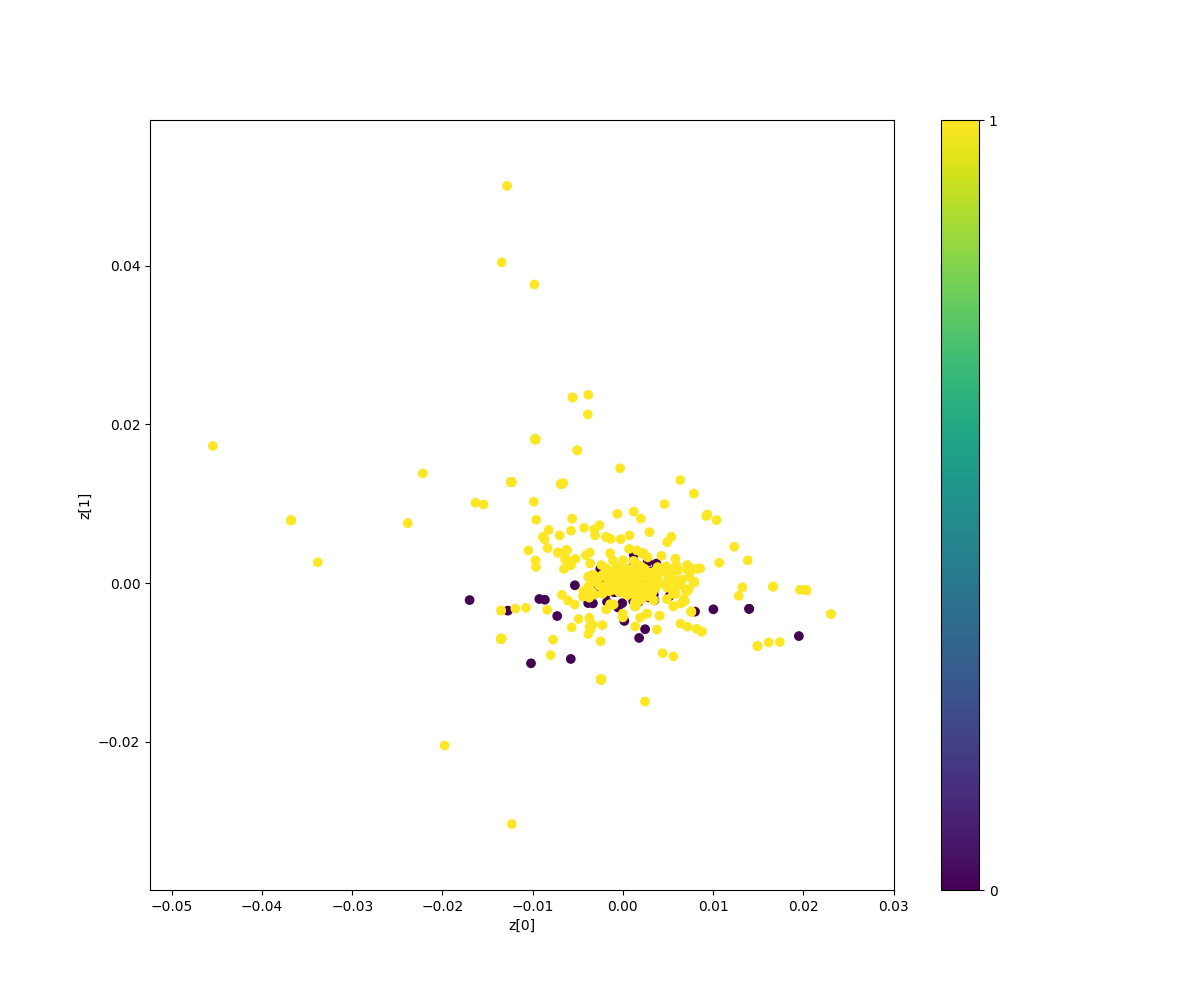}

&
\includegraphics[trim = 20mm 20mm 20mm 20mm, clip, width=0.24\linewidth]{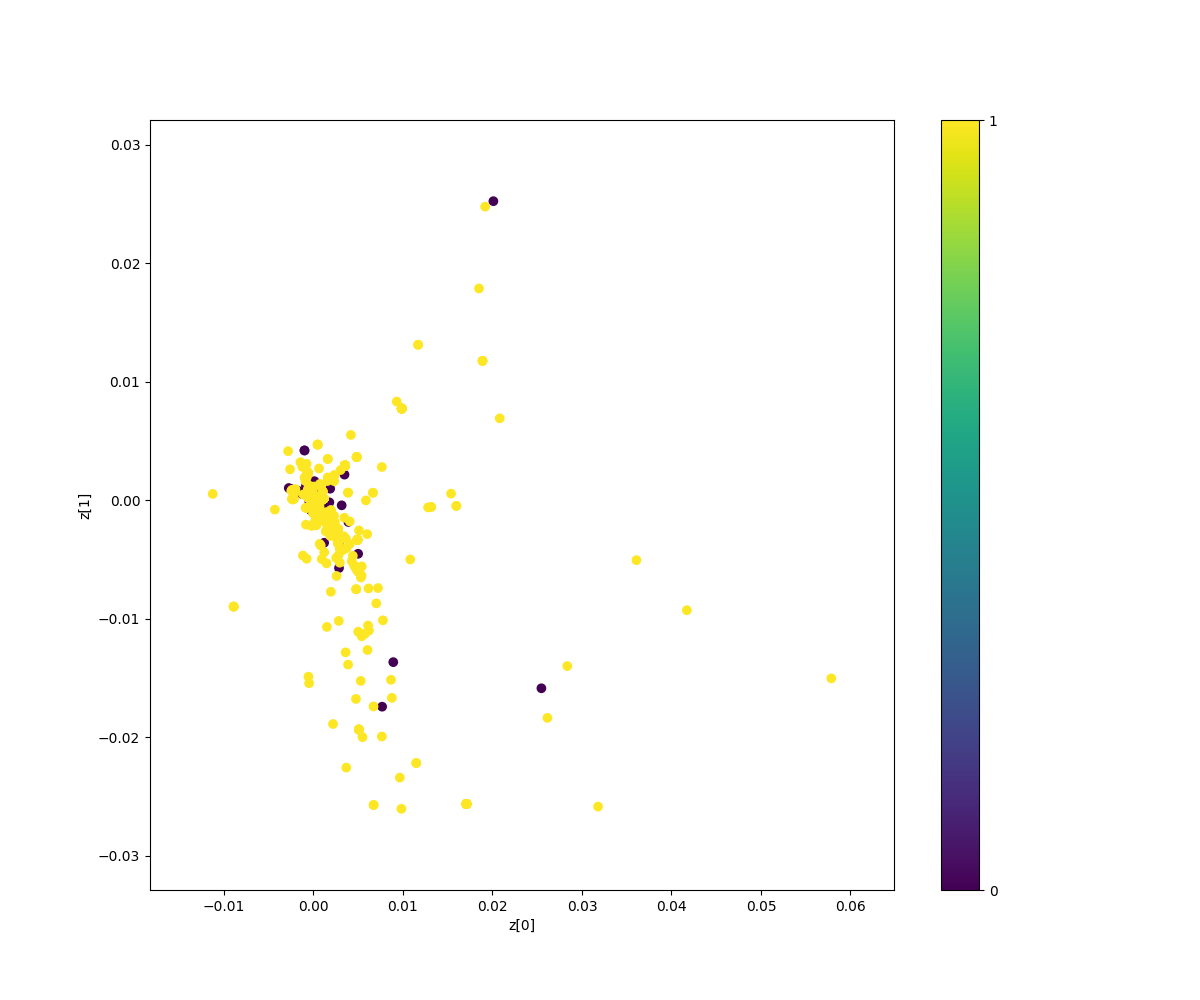}


\end{tabular}
\caption{Latent space visualization - KPCA, VAE-NLL, VAE-MSE for system calls (top) and memory dumps (bottom).}
\label{fig:latent}
\end{figure*}

Starting with system call traces, the main observation is the very similar AUC ROC across the KPCA and all VAE configurations, falling within the 0.691 - 0.708 range, with the maximum score belonging to KPCA. However, in the case of f1 scores KPCA outperforms VAE substantially, obtaining 0.864/0.766/0.99 f1/precision/recall. The similar f1 scores across the VAE configurations, in the ranges of 0.509-0.513/0.577-0.624/0.435-0.455 for f1/recall/precision, justify the 2-dimensional (2D) latent layer topologies (3 \& 6). The NLL/MSE approaches return similar scores. The 3 plots in Figure \ref{fig:latent} (top) show the 2D latent space visualizations for the configurations having a 2-dimensional latent space, and which provide further insight into the obtained scores. In all cases there is substantial overlap in compressed latent spaces, with some visible separability emerging only for outliers. 

Onto process memory dumps (Figure \ref{fig:latent} - bottom), it is surprising to observe a more extensive visible separability across the two classes for KPCA. As for the VAE the situation remains similar to system call traces. These observations translate to the KPCA's f1 shooting up to 0.88 for 0.97/0.8 recall/precision. At least from a KPCA point of view, these results show promise for the Android system service class representation derived from HPROF dumps. Yet, the very similar VAE AUC ROC range, 0.68-0.72, and f1 score range, 0.45-.052, excluding topology 6, indicate that we cannot dismiss VAE as yet. For VAE it is noteworthy that: i) All configurations register a substantial increase in recall (0.81-0.9) however at the cost of a dip in precision (0.34-0.37); ii) Topology 6, that makes use of the MSE loss function, is less accurate, and therefore indicating that there could be cases where fixing $\sigma^{2}_{\hat{x}}$ may not be a good idea. 

\subsection{Discussion}

%


\paragraph{KPCA \& VAE for Android anomalous system call trace detection.} Detection accuracies for both KPCA and VAE using system call traces compare well to those obtained for network anomaly detection \cite{an2015variational}. In that case the registered AUC ROC for KPCA/VAE across the DoS-Probe-R2L-U2R attack categories was 0.590/0.795-0.821/0.944-0.712/0.777-0.712/0.782. The main difference in our case being KPCA outperforming VAE, especially when considering the 0.861 vs 0.513 f1 scores. In the network case the only particularly higher score compared to Android was registered for the Probe category, and which is an exceptionally noisy category.

\paragraph{Process memory dumps.} In combination with the system service call representation, the KPCA detector registers even better effectiveness for HPROF dumps. On the contrary, all VAE configurations have their precision impacted. Yet, the obtained AUC ROC scores do not allow us to commit exclusively to KPCA as of this point, especially when considering the possibility of a custom VAE topology for our application.

\paragraph{Improving app behavior representation.} A compelling idea in this regard is to combine the call tracing and memory dump approaches into a single \emph{online object collection}. The combined approach entails tracing just the \texttt{getSystemService()} API call, and at which point to dump the corresponding service class instance from memory. In doing so, this combined approach addresses the requirement to time memory dumps in a way to coincide with the in-memory presence of the sought-after heap objects.

\subsection{Related work}\label{section:relwork}

Stacking multiple AEs and appending with fully connected layers, forming a deep belief network, provide effective architectures for malware classification \cite{hou2016deep4maldroid,hardy2016dl4md}. For network anomaly detection VAEs give better results than AEs \cite{an2015variational}, with a particular study suggesting that models may be improved further with supervised learning \cite{lopez2017conditional}. In a context where deep learning is under the spotlight, experimentation with kernel methods is still ongoing and yielding promising results \cite{callegari2018improving}. On  other hand, the use of machine learning for memory forensics is still in its early stages of experimentation, with efforts working directly with raw process memory \cite{kumara2017leveraging} also being proposed.

\section{Conclusions \& future work}\label{section:concl}



In this paper we proposed SpotCheck, an on-device anomaly detector for Android malware. Results show that we managed to reproduce the level of effectiveness within an Android anomaly detection context, what previously had been done with VAEs for network anomaly detection. Further still, we showed that HPROF dumps can replace system call traces without impacting effectiveness. Future work will focus on accurate collection of in-memory system service objects, as well as experimentation with fully-supervised VAE topologies. Lastly, we need to close the loop by showing how malware sandboxes can benefit from the identified anomalous execution traces.

\begin{acks}
This work is supported by the LOCARD Project under Grant H2020-SU-SEC-2018-832735.
\end{acks}

\bibliographystyle{ACM-Reference-Format}
\bibliography{mybibliography}


\end{document}